\documentclass[journal, draftcls, one column, 12pt]{IEEEtran}

\usepackage{subfigure}
\usepackage{amsmath}
\usepackage{graphicx}
\usepackage{amssymb}
\usepackage{cases}
\usepackage{color}
\usepackage{cite}
\usepackage{caption}
\usepackage[top=1in, bottom=1in, left=0.7in, right=0.7in]{geometry}

\graphicspath{{./figs/}}

\begin{document}
\title{Wireless Communications with Unmanned Aerial Vehicles: Opportunities and Challenges}
\author{Yong~Zeng, Rui~Zhang, and Teng Joon Lim
\thanks{The authors are with the Department of Electrical and Computer Engineering, National University of Singapore (e-mail: \{elezeng, elezhang, eleltj\}@nus.edu.sg).}
}

\maketitle

\begin{abstract}
Wireless communication systems that include unmanned aerial vehicles (UAVs) promise to provide cost-effective wireless connectivity for devices without infrastructure coverage. Compared to terrestrial communications or those based on high-altitude platforms (HAPs), on-demand wireless systems with low-altitude UAVs are in general faster to deploy, more flexibly re-configured, and are likely to have better  communication channels due to the presence of short-range line-of-sight (LoS) links. However, the utilization of highly mobile and energy-constrained UAVs for wireless communications also introduces many new challenges.  In this article, we provide an overview of UAV-aided wireless communications, by introducing the basic networking architecture and main channel characteristics, highlighting the key design considerations as well as the new opportunities to be exploited.
\end{abstract}

\section{Introduction}
With their high mobility and low cost, unmanned aerial vehicles (UAVs),   also commonly known as drones or remotely piloted aircrafts,  have found a wide range of applications during the past few decades~\cite{616}. Historically, UAVs have been primarily used in the military, mainly deployed in hostile territory to reduce pilot losses.  With the continuous cost reduction and device miniaturization, small UAVs (typically with weight not exceeding 25 kg) are now more easily accessible to the public and thus numerous new applications in civilian and commercial domains have emerged, with typical examples including weather monitoring, forest fire detection, traffic control, cargo transport, emergency search and rescue, communication relaying, etc \cite{614}.
UAVs can be broadly classified into two categories: fixed wing versus rotary wing, each with their own strengths and weaknesses. For example, fixed-wing UAVs usually have high speed and heavy payload, but they must maintain a continuous forward motion to remain aloft, thus are not suitable for stationary applications like close inspection. In contrast, rotary-wing UAVs such as quadcopters, though having limited mobility and payload, are able to move in any direction as well as to stay stationary in the air. Thus, the choice of UAVs critically depends on the applications.

Among the various applications enabled by UASs, the use of UAVs for achieving high-speed wireless communications is expected to play an important role in future communication systems. In fact, UAV-aided wireless communication offers one promising solution to provide wireless connectivity for devices without infrastructure coverage due to e.g., severe shadowing by urban or mountainous terrain, or damage to the communication infrastructure caused by natural disasters \cite{615}.  Note that besides UAVs, one alternative solution for wireless connectivity is via high-altitude platforms (HAPs), such as balloons, which usually operate in the stratosphere that is tens of kilometers above the Earth's surface. HAP-based communications have several advantages over the UAV-based low-altitude platforms (LAPs), such as wider coverage, longer endurance, etc. Thus, HAP is in general preferred for providing reliable wireless coverage for a large geographic area. However, compared to HAP-based communications, or those based on terrestrial or satellite systems,  wireless communications with low-altitude UAVs (typically at an altitude not exceeding several kilometers) also have several important advantages.  First, on-demand  UASs are more cost-effective and can be much more swiftly deployed, which makes them especially suitable for unexpected or limited-duration missions. Besides, with the aid of low-altitude UAVs, short-range line-of-sight (LoS)  communication links can be established in most  scenarios, which potentially leads to significant performance improvement over direct communication between source and destination (if possible) or HAP relaying over long-distance LoS links. In addition, the maneuverability of UAVs offers new opportunities for performance enhancement, through the dynamic adjustment of UAV state to best suit the communication environment. Furthermore, adaptive communications can be jointly designed with UAV mobility control to further improve the communication performance. For example, when a UAV experiences good channels with the ground terminals, besides transmitting with higher rates, it can also lower its speed to sustain the good wireless connectivity to transmit more data to the ground terminals. These evident benefits make UAV-aided wireless communication a promising integral component of future wireless systems, which need to support more diverse applications with orders-of-magnitude capacity improvement over the current systems.  Fig.~\ref{F:ArchitecturesAll} illustrates three typical use cases of UAV-aided wireless communications, which are:

\begin{figure}
\centering
\includegraphics[scale=0.6]{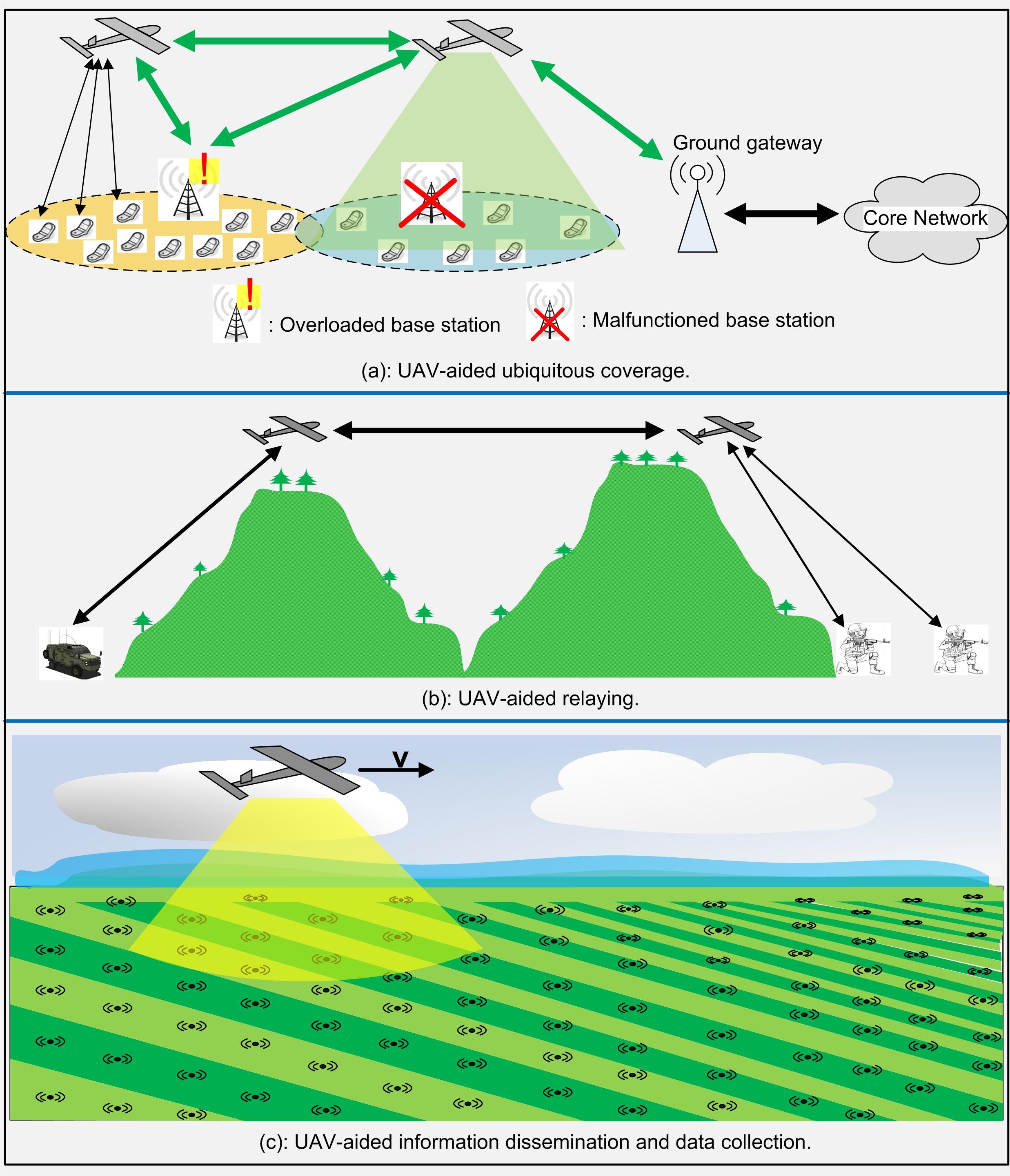}
\caption{Three typical use cases of UAV-aided wireless communications.}\label{F:ArchitecturesAll}
\end{figure}

(a) {\it UAV-aided ubiquitous coverage}, where UAVs are deployed to assist the existing communication infrastructure, if any, in providing seamless wireless coverage within the serving area. Two example scenarios are rapid service recovery after partial or complete infrastructure damage due to natural disasters, and base station offloading in extremely crowded areas, e.g., a stadium in a sports event. Note that the latter case has been identified as one of the five key scenarios that need to be effectively addressed by the fifth generation (5G) wireless systems \cite{613}.

(b) {\it UAV-aided relaying}, where UAVs are deployed to provide  wireless connectivity between two or more distant users or  user groups without reliable direct communication links, e.g., between the frontline and the command center for emergency responses.

(c) {\it UAV-aided information dissemination and data collection}, where UAVs are despatched to disseminate (or collect) delay-tolerant information  to (from) a large number of distributed wireless devices, e.g., wireless sensors in {\it precision agriculture} applications.

Despite  the many promising benefits,  wireless communications with UAVs are also faced with several new design challenges. First, besides the normal communication links as in terrestrial systems, additional control and non-payload communications (CNPC) links with much more stringent latency and security requirements are needed in UASs for supporting   safety-critical functions, such as real-time control, collision and crash avoidance, etc. This calls for more effective resource management and security mechanisms specifically designed for UAV communication systems. Besides, the high mobility environment of UASs generally results in highly dynamic network topologies, which are usually sparsely and intermittently connected \cite{618}. As a result, effective multi-UAV coordination, or UAV swarm operations,  need to be designed for ensuring reliable network connectivity~\cite{617}. At the same time, new communication protocols need to be designed taking into account the possibility of sparse and intermittent network connectivity. Another main challenge stems from the size, weight, and power (SWAP) constraints of UAVs, which could limit their communication, computation, and endurance capabilities.  To tackle such issues, energy-aware UAV deployment and operation  mechanisms are needed  for intelligent energy usage and replenishment.  Moreover, due to the mobility of UAVs as well as the lack of fixed backhual links and centralized control, interference coordination among the neighboring cells with UAV-enabled aerial base stations is more challenging than in terrestrial cellular systems. Thus, effective interference management techniques specifically designed for UAV-aided cellular coverage are needed.

The objective of this article is to give an overview of UAV-aided wireless communications.  The basic networking architecture,  main channel characteristics and design considerations, as well as the key performance enhancing techniques that exploit the UAV's mobility will be presented.

\begin{figure}
\centering
\includegraphics[scale=0.5]{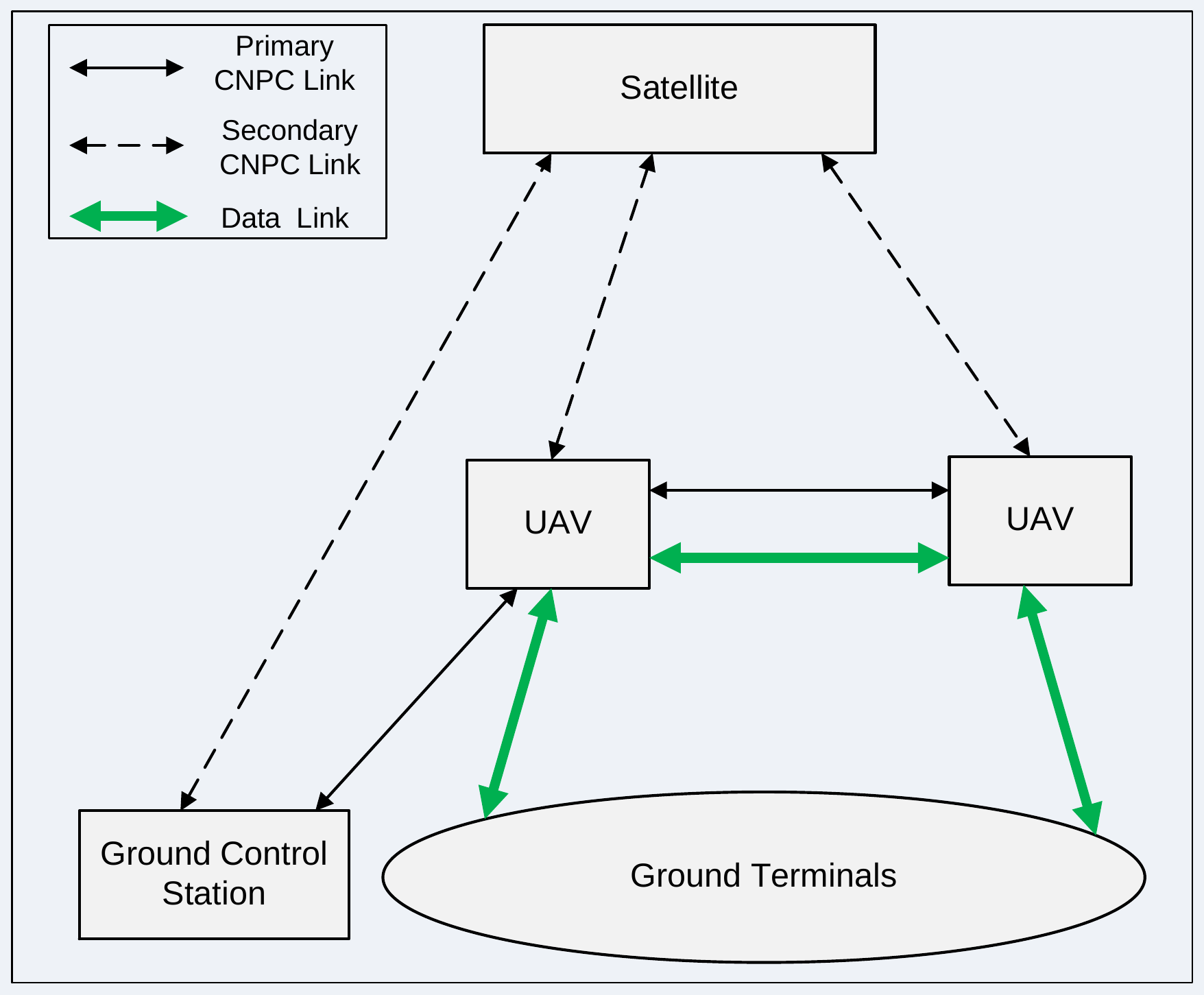}
\caption{Basic networking architecture of UAV-aided wireless communications.}\label{F:NetworkModel}
\end{figure}

\section{Basic Networking Architecture}
Fig.~\ref{F:NetworkModel} shows the generic networking architecture of wireless communications with UAVs, which consists of two basic types of communication links, namely the CNPC link and the data link.

\subsection{Control and Non-Payload Communications Link}
The CNPC links are essential to ensure the safe operation of all UASs. Highly reliable, low-latency, and secure two-way communications, usually with low data rate requirement, must be supported by these links for exchanging safety-critical information among UAVs, as well as between the UAV and ground control stations (GCS), e.g., dedicated mobile terminals mounted on ground vehicles. The main CNPC information flow can be broadly categorized into three types: i) command and control from GCS to UAVs; ii) aircraft status report from UAVs to ground; iii) sense-and-avoid information among UAVs. Even for autonomous UAVs, which are able to accomplish missions relying on onboard computers without real-time human control, the CNPC links are also necessary in case emergency human intervention is needed. Not shown in Fig.~\ref{F:NetworkModel} are the air traffic control (ATC) links, which are  necessary only when the UAVs are within a controlled airspace, e.g., near an airport.

Due to the critical functions to be supported, CNPC links should in general operate in protected spectrum. Currently two such bands have been allocated, namely the L-band (960-977MHz) and the C-band (5030-5091MHz) \cite{621}. Furthermore, although the direct links between GCS and UAVs (primary CNPC links) are always preferred for delay reasons,  secondary CNPC links via satellite could also be exploited as a backup to enhance reliability and robustness. Another key requirement for CNPC links is the superior high security. In particular, effective security mechanisms should be employed to avoid the so-called {\it ghost control} scenario, a potentially catastrophic situation in which the UAVs are controlled by unauthorized agents via  spoofed control or navigation signals. Therefore, powerful authentication techniques, possibly complemented by the emerging physical layer security techniques,  should be applied for CNPC links.

\subsection{Data Link}
The data links, on the other hand, aim to support mission-related communications  for the ground terminals, which, depending on the application scenarios, may include terrestrial base stations (BSs), mobile terminals, gateway nodes, wireless sensors, etc. Taking the UAV-aided ubiquitous coverage shown in Fig~\ref{F:ArchitecturesAll}(a) as an example, the data links maintained by the UAVs need to support the following communication modes: i) direct mobile-UAV communication as for BS offloading or during complete BS malfunction; ii) UAV-BS and UAV-gateway wireless backhaul; iii) UAV-UAV wireless backhaul. The capacity requirement for these data links critically depends on the applications, possibly ranging from several kbps in UAV-sensor links  to dozens of Gbps in UAV-gateway wireless backhaul. Compared to CNPC links, the data links usually have higher tolerance in terms of latency and security requirements. In terms of spectrum,  the UAV data links could reuse the existing band assigned for the particular applications to be supported, e.g., the LTE band while assisting cellular coverage, or dedicated new spectrum could be allocated for enhanced performance, e.g., using millimeter wave (mmWave) band for high capacity UAV-UAV wireless backhaul \cite{569}.

\section{Channel Characteristics}
Both CNPC and data links in  UAV-aided communications  consist of two types of channels, namely UAV-ground and UAV-UAV channels, which exhibit several unique characteristics as compared to the extensively studied terrestrial communication channels.

\subsection{UAV-Ground Channel}\label{sec:UAVGroundChannel}
While the  air-ground channels for aeronautical applications  with piloted aircrafts are well understood,  systematic measurements and modeling of UAV-ground channels are still ongoing\cite{621,622}. Unlike piloted aircraft systems, where the ground sites are usually in open areas with tall antenna towers, the UAV-ground channels for UASs are more complicated due to the more complex operation environment. While LoS links are expected for such channels in most scenarios, they could also be occasionally blocked by obstacles such as terrain, buildings, or the airframe itself. In particular, recent measurements have shown that the UAV-ground channels could suffer from severe airframe shadowing with a duration up to dozens of seconds during aircraft maneuvering  \cite{622}, which needs to be taken into account for mission-critical operations.  For low-altitude UAVs, the UAV-ground channels may also constitute a number of multi-path components due to reflection, scattering, and diffraction by mountains, ground surface, foliage, etc. For UAVs operating over desert or sea, the two-ray model has been mostly used due to the dominance of the LoS and the surface reflection components. Another widely used model is the stochastic Rician fading model, which consists of a deterministic LoS component, and a random scattered component with certain statistical distributions.   Depending on the environment surrounding the ground terminals as well as the frequency used, the UAV-ground channels exhibit widely varying Rician factors, i.e., the power ratio between the LoS and the scattered components,  with typical values around 15 dB for L-band and 28 dB for C-band in hilly terrain \cite{621}.

\subsection{UAV-UAV Channel}
The UAV-UAV channels are mainly dominated by the LoS component. Although there may exist limited multipath fading due to ground reflections, its impact is minimal as compared to that experienced in UAV-ground or ground-ground channels. In addition,  the UAV-UAV channels may have even higher Doppler frequencies than the UAV-ground counterparts, due to the potentially large relative velocity between UAVs. Such channel characteristics have direct implications on spectrum allocation for UAV-UAV links. On one hand, the dominance of LoS links may suggest that the emerging mmWave communications could be employed to achieve high-capacity UAV-UAV wireless backhaul. On the other hand, the high relative velocity between UAVs coupled with the higher frequency in the mmWave band could lead to excessive Doppler shift. More in-depth studies are needed to find out the most suitable technology to use in UAV-UAV links, given their unique channel characteristics.

\section{Main Design Considerations}
This section presents the main design considerations specifically for wireless communications with UAVs. The following three aspects are discussed: UAV path planning, energy-aware deployment and operation, and multiple-input multiple-output (MIMO) communications in UASs.
\subsection{UAV Deployment and Path Planning}
One important design aspect of UASs is UAV path planning \cite{619,620}. For UAV-aided communications in particular, appropriate path planning may significantly shorten the communication distance  and thus is crucial for high-capacity performance. Unfortunately, finding the optimal flying path for UAV is a challenging task in general. On one hand, UAV path optimization problems essentially involve an infinite number of variables due to the continuous UAV trajectory to be determined. On the other hand, the problems are also usually subject to a variety of practical constraints, e.g.,  connectivity, fuel limitation, collision and terrain avoidance,  many of which are time-varying in nature and are difficult to  model accurately. One useful method for UAV path planning is to approximate the UAV dynamics by a discrete-time state space, with the state vector typically consisting of the position and velocity in a three-dimensional (3D) coordinate system. The UAV trajectory is then given by the sequence of states, which are subject to finite transition constraints to reflect the practical UAV mobility limitations. Many of the resulting problems with such an approximation belong to the class of mixed integer linear programming (MILP)  \cite{620}, which can be solved with well-developed software packages.

Intuitively, the optimal UAV flight path critically depends on the application scenarios. For instance, for UAV-aided cellular coverage as shown in Fig.~\ref{F:ArchitecturesAll}(a), it is evident that more than one  UAVs should be jointly deployed above the serving areas  to cooperatively achieve real-time communications with ground users; whereas for UAV-aided information dissemination or collection for delay-tolerant data, as shown in Fig.~\ref{F:ArchitecturesAll}(c), it could be sufficient to despatch one single UAV to fly over the area to communicate with the ground nodes sequentially. Furthermore, for the cellular coverage application, one option is to employ rotary-wing UAVs that hover above the coverage area,  serving as static aerial base stations. In this case, no dedicated path planning is needed. Instead, the main design problems for UAV deployment usually involve finding the optimal UAV separations as well as their hovering altitude to achieve maximum coverage. Note that for a typical urban environment, there in general exists an optimal UAV altitude in terms of coverage maximization, which is due to the following non-trivial tradeoff: While increasing UAV altitude will lead to higher free space path loss, it also increases the possibility of having LoS links with the ground terminals. Such a tradeoff has been characterized in \cite{642,643}, based on which the optimal UAV altitude has been obtained.

\subsection{Energy-Aware Deployment and Operation}
The performance and operational duration of a UAS is fundamentally constrained by the limited onboard energy. Although powerplant and energy-storage technologies have advanced dramatically  over the past few decades, limited energy availability still severely hampers UAV endurance. From the operational perspective, this problem can be addressed through two approaches. First, effective {\it energy-aware deployment} mechanisms are needed for timely onboard energy replenishment, yet without noticeable interruption of the communication services supported. Second, {\it energy-efficient operation}  through smart energy management is required, i.e., accomplishing the missions with minimum energy consumption.

In terms of  energy-aware deployment, one effective approach is to exploit the inter-UAV cooperation to enable sequential energy replenishment. For instance, at any one time, only one UAV is scheduled  to leave the serving area for energy replenishment, during which the service gap is temporarily filled by neighboring UAVs via e.g., increasing the transmission power and/or adjusting the aircraft positions. This energy replenishment scheduling can be matched to the dynamic load patterns that need to be supported by the UAVs. For instance, it might be preferred to schedule  energy replenishment only when low data traffic is expected, e.g., during night time for the cellular coverage application. Note that apart from the commonly used energy sources such as electric batteries or liquid fuels, there has been increasing interest in powering UAVs by solar energy or dedicated wireless energy transfer technology via e.g. laser beams.\footnote{See the company website http://lasermotive.com/ for more details.}

Energy-efficient operation, on the other hand, aims to reduce unnecessary energy consumption by the UAVs. As the main energy usage of UAVs is to support either aircraft propulsion or wireless communications,  energy-efficient operation schemes can be broadly classified into two categories. The first one is {\it energy-efficient mobility}, for which the movement of the UAVs should be carefully controlled by taking into account the energy consumption associated with every maneuver. For instance, unnecessary aircraft maneuvering or ascending should be avoided since they are generally quite energy-intensive. Energy-efficient mobility schemes can  usually be designed with path planning optimization, by using appropriate energy consumption models as a function of UAV speed, acceleration, altitude, etc. The other category of energy-efficient operation is {\it energy-efficient communication}, which aims to satisfy the communication requirement with the minimum energy expenditure on  communication-related functions, such as communication circuits, signal transmission, etc. To this end, one common approach is to optimize the communication strategies to maximize the {\it energy efficiency} (EE) in bits/Joule, i.e., the number of successfully communicated data bits per unit energy consumption. Note that while energy-efficient communication has been extensively studied for terrestrial communications,  its systematic investigation for UAV communication systems is still under-developed.

\subsection{MIMO for UAV-Aided Communications}
Although MIMO technology has been extensively implemented in terrestrial  communication systems due to its high spectral efficiency and superior diversity performance, its application in UASs is still hindered by several factors. First, the lack of rich scattering in UAS environment considerably  limits the spatial multiplexing gain of MIMO, which usually leads to only marginal rate improvement over single-antenna systems. Besides, the high signal processing  complexity as well as the hardware and power consumption costs make it  quite costly to employ multiple antennas in UAVs due to the SWAP limitations. Furthermore, MIMO systems rely on  accurate channel state information (CSI) for best performance. However, this is practically difficult to achieve in a highly dynamic environment, therefore further limiting the practical MIMO gain in UASs.

Despite the above challenges, some recent results still show a great potential for MIMO technology in UASs. In particular, in contrast to the common conception that spatial multiplexing gain is fundamentally limited by the number of signal paths, it has been found that high spatial multiplexing gain may also be attainable even in LoS channels, by carefully designing the antenna separation with respect to carrier wavelength and link distance \cite{624}, though this usually requires large antenna separation, high carrier frequency, and short communication range. Alternatively, a more practical way to reap the multiplexing gain in poor scattering environment is to leverage multi-user MIMO, by simultaneously serving a number of sufficiently separated ground terminals with angular separations exceeding the angular resolution of the antenna array installed on the UAVs. In this case, the signals for different terminals are distinguishable by the UAV array, and thus restores the MIMO spatial multiplexing gain.  Another way of utilizing MIMO in UASs is through mmWave communications, for which the MIMO array gain, instead of the spatial multiplexing gain, is more critical due to the large available bandwidth as well as the high signal attenuation. However, due to the high mobility of UAVs, it would be quite challenging to achieve transmitter/receiver beam alignment for directional mmWave communications, an issue that needs to be properly addressed before mmWave MIMO could be practically employed in UAV systems.

\section{Communications with UAV Controlled Mobility}
The high mobility of UAVs offers unique opportunities for performance improvement in UAV-aided communications. In this section, we discuss two key techniques for wireless communications with UAV controlled mobility, which are UAV-enabled mobile relaying and device-to-device (D2D)-enhanced UAV information dissemination.

\subsection{UAV-Enabled Mobile Relaying}
Relaying is an extensively studied technique in terrestrial  communication systems for throughput/reliability improvement as well as range extension. Due to the practical  constraints such as limited mobility and wired backhauls, most relays in terrestrial systems are deployed  in fixed locations, which we term as {\it static relaying}. To further exploit the UAV controlled mobility, we present in this subsection a UAV-enabled {\it mobile relaying} strategy, which works particularly well for delay-tolerant applications.

\begin{figure}
\centering
\subfigure[A schematic of the UAV-enabled mobile relaying]{
\includegraphics[scale=0.6]{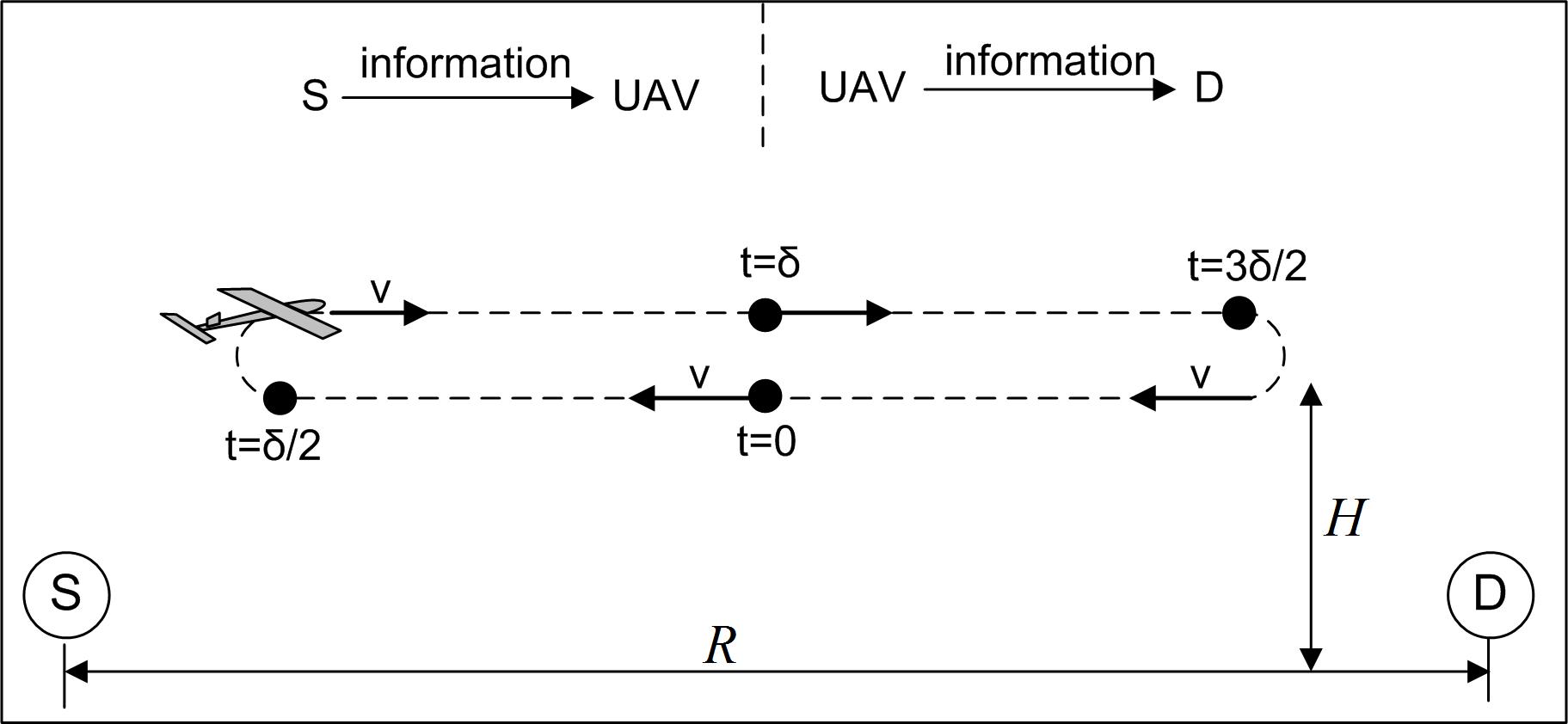}
} \vspace{-1ex} \\
\subfigure[Path loss with static versus mobile relaying]{
\includegraphics[scale=0.7]{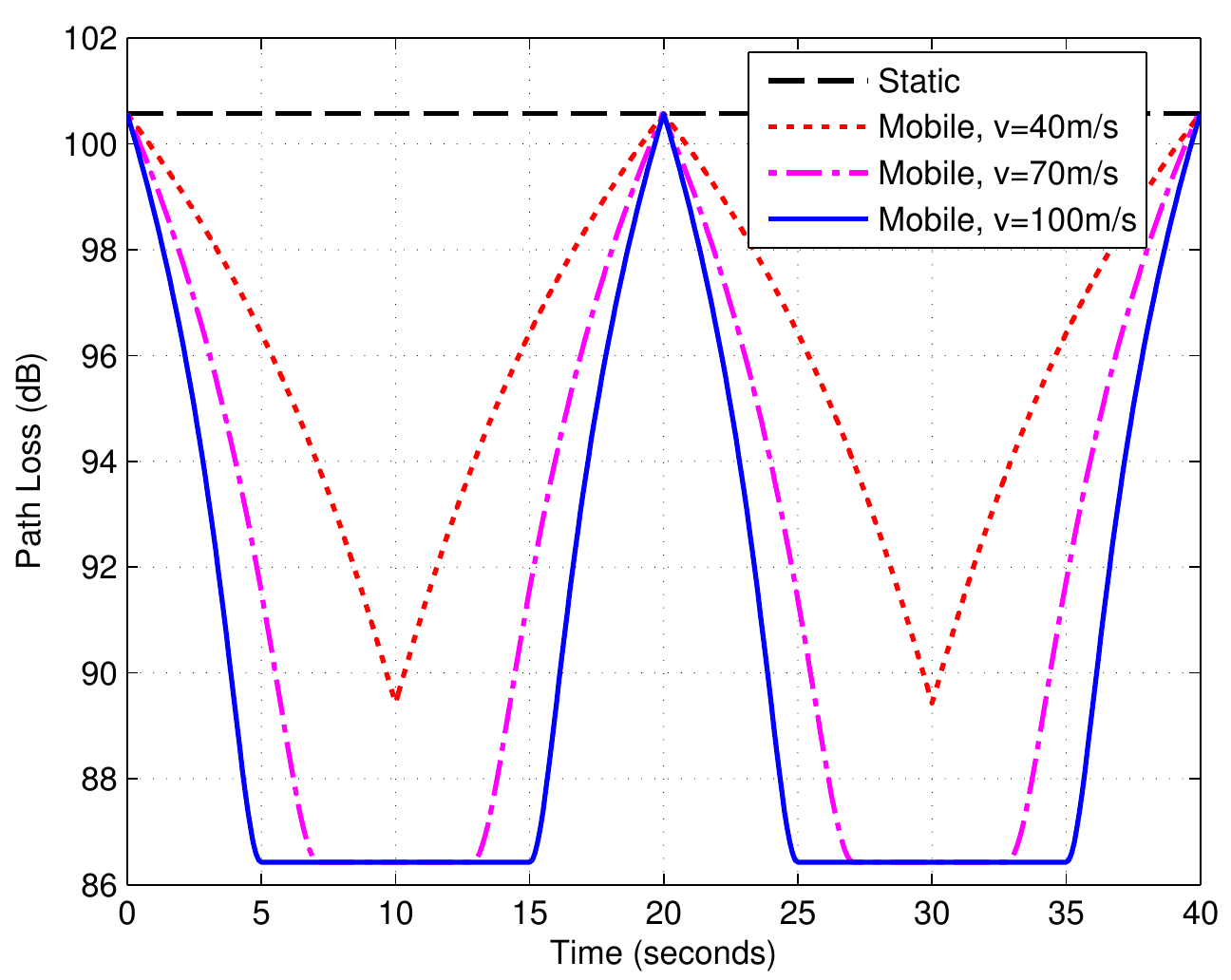}
} \vspace{-2ex}
\caption{UAV-enabled mobile relaying and the corresponding path loss of the communication links.}%
\label{F:MobileRelay}
\end{figure}

 With mobile relaying, the UAV flies continuously between the source and destination aiming to reduce the link distances during both UAV information reception and relaying phases. For example, with half-duplex decode-and-forward (DF) mobile relaying, each  relaying cycle consists of two phases each with duration $\delta$ seconds, where $\delta$ is determined by the maximum tolerable delay. As illustrated in Fig.~\ref{F:MobileRelay}(a), the first phase corresponds to UAV information reception, where it keeps receiving and decoding the information sent from the source and stores in its data buffer. Concurrently, starting from the initial position at the middle point between the source and destination, the UAV first flies towards the source at a maximum possible speed $v$, and then flies back timely so that it returns to the initial position at the end of the first phase ($t=\delta$). Note that if $v$ and/or $\delta$ is sufficiently large, the UAV will have time to hover above the source before returning so as to enjoy the best channel for data reception. In the second phase starting from $t=\delta$, the UAV sends the data in its buffer to the destination. This is accompanied by a symmetric UAV movement, where it first flies towards the destination, hovers above the nearest location to the destination if time allows, and then returns to the initial position  at the end of the cycle ($t=2\delta$). It is evident that compared to static relaying with the fixed UAV location at the same initial position, the proposed mobile relaying strategy always enjoys a shorter link distance (or better average channel) in each of the two phases of information reception and relaying. This is illustrated in Fig.~\ref{F:MobileRelay}(b) with $\delta=20$ seconds under different UAV velocity and a constant height $H=100$ m. The carrier frequency is $5$ GHz and the source and destination are assumed to be separated by $R=1$ km.  It is observed from Fig.~\ref{F:MobileRelay}(b) that with higher UAV speed limitation, mobile relaying enjoys larger link gains over static relaying.  In particular, for sufficiently large UAV maximum speed, e.g., $v=100$m/s, the UAV would be able to stay stationary above the source and destination each for about $10$ seconds, during which the path loss remains at a constant value that is about $14$ dB lower than that of the static relaying.

\begin{figure}
\centering
\includegraphics[scale=0.6]{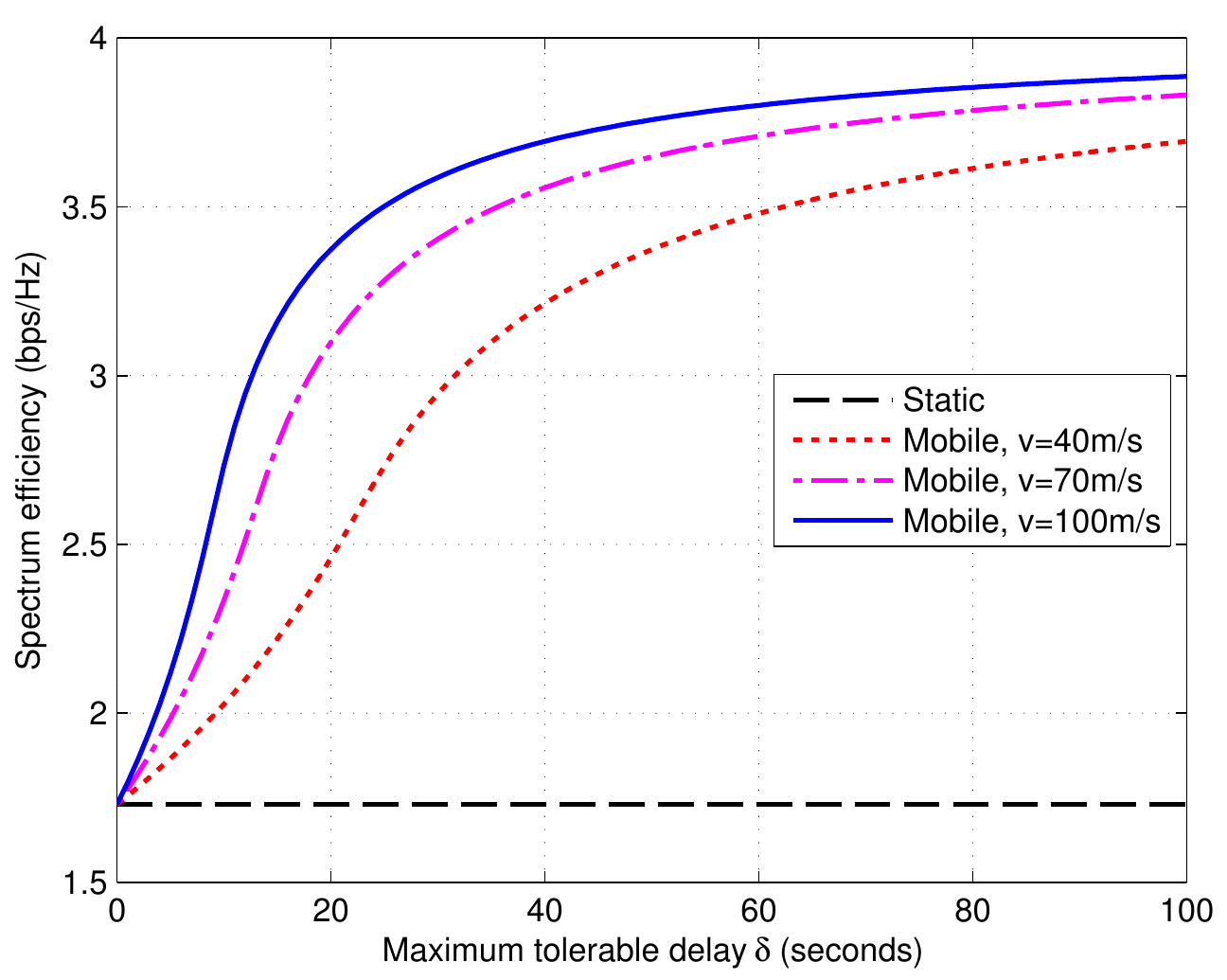}
\caption{Spectrum efficiency versus maximum tolerable delay with mobile versus static relaying.\vspace{-3ex}}\label{F:ThroughputVSDelay}
\end{figure}

By employing adaptive rate transmission, the proposed mobile relaying strategy can achieve significant throughput improvement over the conventional  static relaying. This is illustrated in Fig.~\ref{F:ThroughputVSDelay}, where the end-to-end spectrum efficiency in bps/Hz is plotted against the maximum tolerable delay $\delta$ for different UAV velocity. Both the source and the UAV are assumed to transmit with a constant power $P$, with $P$ setting to a value so that the average received signal-to-noise ratio (SNR) at the UAV for the static relaying is $10$ dB. Note that the direct link between source and destination is assumed to be blocked and thus ignored. For simplicity, we assume that the Doppler effect due to UAV's mobility has been well compensated. It is observed that for sufficiently high delay tolerance $\delta$, the mobile relaying strategy achieves a throughput more than twice of that by static relaying. Furthermore, for any fixed $\delta$, larger throughput is achieved  for higher UAV velocity, which is as expected.

Note that an alternative strategy of  mobile relaying is known as {\it data ferrying} or {\it load-carry-and-delivery} \cite{618}. With this strategy, the UAV ``loads'' the data from the source as it reaches the nearest possible location from the source, flies towards the destination with the loaded data until it reaches the nearest possible location to the destination, and then delivers the data to the destination. As data ferrying has less communication time than the proposed mobile relaying, its achievable throughput is expected to be smaller, especially for cases with low UAV speed and/or stringent delay requirement.  Furthermore, in the above discussions, a data buffer with sufficiently large buffer size is assumed at the UAV. In general, there exists a trade-off between on-board buffer size and achievable throughput in the mobile relaying design.

\subsection{D2D-Enhanced UAV Information Dissemination}
D2D communication is an effective technique for capacity improvement in terrestrial communication systems \cite{628}. The main idea is to offload the BS by enabling direct communications between nearby mobile terminals. For UAV-aided communication systems, D2D communication is expected to play an  important role by providing the additional benefits such as UAV energy saving, lower capacity requirement for UAV wireless backhaul, etc. Many existing D2D techniques for terrestrial communication systems, such as those on interference mitigation and spectrum sharing, can be directly applied in UAV-aided communications, especially in the scenario to support ubiquitous cellular coverage as shown in Fig.~\ref{F:ArchitecturesAll}(a). On the other hand, new D2D communication techniques could  be devised by exploiting the unique characteristics of UAV-aided communications. In the following, we present one such technique, termed {\it D2D-enhanced UAV information dissemination}, which aims to achieve efficient information dissemination to a large number of ground nodes by exploiting both D2D communications and the UAV  mobility.

\begin{figure}
\centering
\includegraphics[scale=0.6]{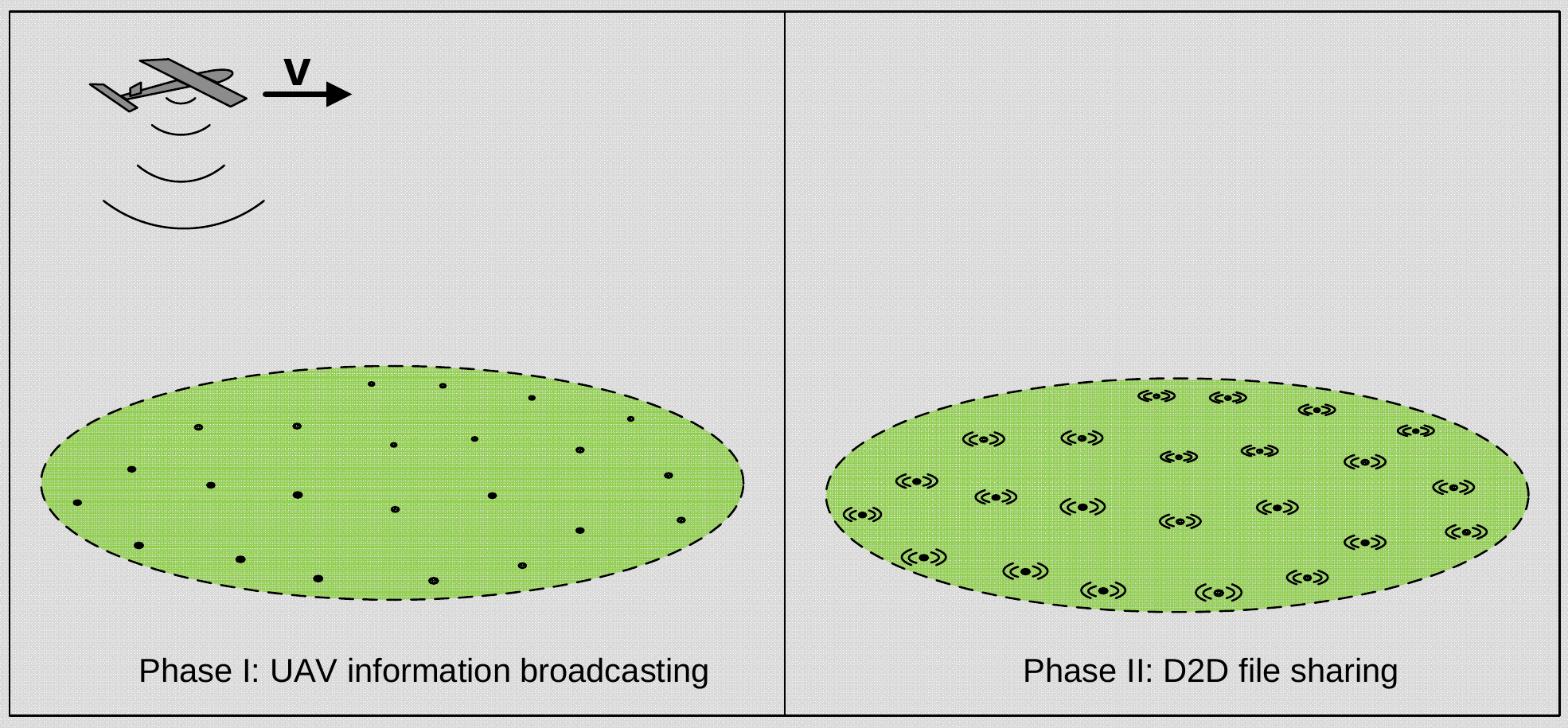}
\caption{The two-phase protocol of D2D-enhanced UAV information dissemination.\vspace{-3ex}}\label{F:D2D}
\end{figure}

As illustrated in Fig.~\ref{F:ArchitecturesAll}(c), we consider the scenario where one UAV flies over a certain area to distribute a common file to a large number of ground nodes. One simple approach to achieve this is by letting the UAV repeatedly transmit  the same file as it flies over different ground nodes, until all of them successfully receive the file. It is not difficult to see that such a scheme requires substantial  UAV retransmissions, and its performance is essentially limited by the ground terminals which experience the weakest channel conditions with the UAV. The D2D-enhanced information dissemination scheme can effectively solve this problem with a two-phase protocol, as illustrated in Fig.~\ref{F:D2D}. In the first phase, the UAV broadcasts the appropriately coded file to the ground nodes as it flies over them. Since each node has only limited wireless connectivity with the UAV, it is very likely that it can only successfully receive a fraction of the file, where different portions of the file are received by different nodes. In the second phase, the ground nodes exchange their respectively received data via D2D communications, until all the nodes receive a sufficient number of packets to successfully decode the file. This scheme significantly reduces the number of  UAV retransmissions and as a result the total flying time of the UAV, which saves its energy and is particularly useful for small UAVs with limited onboard energy. Notice that if the ground nodes are distributed over a wide geographical area, efficient node clustering algorithms can be applied, to improve the file sharing performance of short-range D2D communications within each cluster. The joint optimization of the  UAV path planning, coding, node clustering, as well as D2D file sharing for this scenario is an important problem  for future research.

\section{Conclusions}
In this article, we have provided an overview on UAV-aided wireless communications with the help of three use cases: UAV-aided ubiquitous coverage, UAV-aided relaying, and UAV-aided information dissemination. The basic networking architecture and main channel characteristics were introduced. Furthermore, the key design considerations for UAV communications were also discussed. Lastly, we highlighted two key performance enhancing techniques  by utilizing the UAV controlled mobility, including UAV-enabled mobile relaying and D2D-enhanced UAV information dissemination. It is hoped that the challenges and opportunities described in this article will help pave the way for researchers to design and build UAV-enhanced wireless communications systems in the future.

\end{document}